\documentclass[10pt, a4paper,english]{elsarticle}

\usepackage[utf8]{inputenc}
\usepackage[T1]{fontenc}
\usepackage{babel}
\usepackage{amsmath}
\usepackage{amssymb}
\usepackage{graphicx}
\usepackage{fancyhdr}
\usepackage{url}
\usepackage[numbers]{natbib}
\usepackage{nameref}

\usepackage[font=small]{caption}

\usepackage{color}

\usepackage{float}  
\restylefloat{table}  

\journal{SoftwareX}
\usepackage{hyperref}

\begin{document}

\begin{frontmatter}  

\title{Open data from the first and second observing runs of Advanced LIGO and Advanced Virgo}

\author{
R.~Abbott,$^{1}$  
T.~D.~Abbott,$^{2}$  
S.~Abraham,$^{3}$  
F.~Acernese,$^{4,5}$ 
K.~Ackley,$^{6}$  
C.~Adams,$^{7}$  
R.~X.~Adhikari,$^{1}$  
V.~B.~Adya,$^{8}$  
C.~Affeldt,$^{9,10}$  
M.~Agathos,$^{11,12}$
% ...
}
\author{(The LIGO Scientific Collaboration and the Virgo Collaboration)}

\address{
$^{1}$LIGO, California Institute of Technology, Pasadena, CA 91125, USA 
 
$^{2}$Louisiana State University, Baton Rouge, LA 70803, USA 
 
$^{3}$Inter-University Centre for Astronomy and Astrophysics, Pune 411007, India 
 
$^{4}$Dipartimento di Farmacia, Universit\`a di Salerno, I-84084 Fisciano, Salerno, Italy 
 
$^{5}$INFN, Sezione di Napoli, Complesso Universitario di Monte S.Angelo, I-80126 Napoli, Italy 
 
$^{6}$OzGrav, School of Physics \& Astronomy, Monash University, Clayton 3800, Victoria, Australia 
 
$^{7}$LIGO Livingston Observatory, Livingston, LA 70754, USA 
 
$^{8}$OzGrav, Australian National University, Canberra, Australian Capital Territory 0200, Australia 
 
$^{9}$Max Planck Institute for Gravitational Physics (Albert Einstein Institute), D-30167 Hannover, Germany 
 
$^{10}$Leibniz Universit\"at Hannover, D-30167 Hannover, Germany 
 
$^{11}$Theoretisch-Physikalisches Institut, Friedrich-Schiller-Universit\"at Jena, D-07743 Jena, Germany 
 
$^{12}$University of Cambridge, Cambridge CB2 1TN, UK 
 
% ...
}

\begin{abstract}
Advanced LIGO and Advanced Virgo are monitoring the sky and collecting gravitational-wave strain data
with sufficient sensitivity to detect signals routinely.
In this paper we describe the data recorded by these instruments during their first and second observing runs.
The main data products are gravitational-wave strain time series sampled at 16384~Hz.
The datasets that include this strain measurement can be freely accessed through the Gravitational Wave Open Science Center at \href{http://gw-openscience.org}{http://gw-openscience.org}, together with data-quality information essential for the analysis of LIGO and Virgo data, 
documentation, tutorials, and supporting software.
\end{abstract}

\end{frontmatter}

\vspace{0.5cm}
\noindent Code metadata
\begin{center}
\begin{tabular}{p{6.5cm}p{6.5cm}}
\hline 
Current data version & O1 V1 and O2 R1 \\
Permanent link to code/repository used for this data version & \url{https://doi.org/10.7935/K57P8W9D} and \url{https://doi.org/10.7935/CA75-FM95}\\
 Legal Code License  & Creative Commons Attribution International Public License 4.0\\
 Code versioning system used & NA\\
 Software code languages, tools, and services used & Python, Django\\
 Compilation requirements, operating environments \& dependencies & Unix, Linux, Mac, Windows\\
 Link to documentation/manual & \url{https://www.gw-openscience.org/O1} and \url{https://www.gw-openscience.org/O2}\\
 Support email for questions & \url{mailto:gwosc@igwn.org}\\
\hline 
\end{tabular} 
\end{center}

\section{Motivation and significance}
Gravitational   waves (GWs) are transverse waves in the spacetime metric that travel at the speed of light. They are generated by accelerated masses and more precisely, to lowest order, by time changes of the mass quadrupole~\cite{Maggiore}, such as in the orbital motion of a binary system of compact stars.
GWs were predicted in 1916 by Albert Einstein after the final formulation of the field equations of general relativity~\cite{Einstein:1916cc,Einstein:1918btx}. They were first observed directly in
2015~\cite{abbott16:_obser} by the Laser Interferometer Gravitational-Wave Observatory (LIGO)~\cite{aasi15:_advan_ligo} during its first observing run (O1), which took place from September 12, 2015 to January 19, 2016.

After an upgrade of the detectors, the second observing run (O2) took place from November 30, 2016 to August 25, 2017. Advanced Virgo~\cite{acernese15:_advan_virgo} joined this observing run on August 1, 2017. 
On April 1, 2019, Advanced LIGO and Advanced Virgo initiated their third observing run (O3), lasting almost one year~\cite{observing_scenario}. 
The analysis of O1 and O2 data produced 11 confident detections (10 binary black hole mergers~\cite{abbott16:_obser,GW150914,GW151226,CatalogueO1,GW170104,GW170814} and 1 binary neutron star merger~\cite{GW170817}) and 14
marginal triggers, collected and described in the Gravitational Wave Transient Catalog (GWTC-1)~\cite{LIGOScientific:2018mvr}. 

Notable events in this catalog are the first observed event GW150914~\cite{abbott16:_obser}, the first three-detector event GW170814~\cite{GW170814} and the binary neutron star (BNS) coalescence GW170817~\cite{GW170817}.
This latter event is the first case where gravitational and electromagnetic waves have been observed from a single source~\cite{GW170817mma} offering a unique description of the physical processes at play during and after
the merger of two neutron stars.

The main data product of the LIGO and Virgo detectors is a time series containing the measure of the strain, which will be described more in detail in the section \ref{sec:methods}.  
The LIGO Scientific Collaboration and the Virgo Collaboration (LVC) release their calibrated strain data to researchers outside the LVC and to a broader public that includes amateur scientists, students, etc.
The roadmap for these data releases is described in the LIGO Data Management Plan~\cite{LIGO-DMP} and in the Memorandum of Understanding between Virgo and LIGO~\cite{LIGOVirgo-MOU} (Attachment~A, Sec. 2.9). 
Two types of data release are foreseen. When GW events are discovered and published individually or in a catalog, the LVC releases short segments of GW strain data around the time of the GW events, as in the case of
GWTC-1~\cite{gwosc-GWTC-1}. In addition, a release of the strain recorded during the entire observation run occurs after a proprietary period of internal use, necessary also to validate and calibrate the data.

The cleaned, calibrated GW strain data related to both the O1 and O2 runs were released in January 2018~\cite{gwosc-O1} and in February 2019~\cite{gwosc-O2}, respectively. 
The release of the strain data for the first block of six months of O3 is currently scheduled for April 2021, and November 2021 for the second 6-month block. 

This article focuses on the already-released data from the O1 and O2 runs.
Public access to these data along with extensive documentation and usage instructions are provided through the Gravitational Wave Open Science Center (GWOSC)~\cite{Vallisneri:2014vxa} at
\url{http://gw-openscience.org}. GWOSC also provides online tools for finding and viewing data, usage guidelines and tutorials. 
We summarize this information, and include a comprehensive bibliography describing the detectors, the data collection and calibration, the noise characterization and software packages for data analysis.

To date more than 200 scientific articles have been written using the data from the GWOSC website. 
These analyses confirm, complement and extend the results published by the LVC, demonstrating the impact on the scientific community of the GW data releases.
The covered topics span from alternative methods to search for gravitational wave events, some leading to new detections,
e.g.~\cite{ref31,ref32,ref38,ref39,ref40,2-OGC,2BBH,PhysRevD.93.022002,PhysRevD.93.082005,GREEN2018312,PhysRevD.100.124022,10.1093/ptep/ptz043,PhysRevD.99.124035}, to reassessed estimations of the event parameters,
e.g.~\cite{ref34,ref37,gerosa_vitale_haster_chatziioannou_zimmerman_2017,PhysRevD.100.104004,PhysRevD.99.124005,PhysRevD.100.104015,PhysRevD.101.103004,PhysRevD.100.084041}, studies on matter effects for the binary neutron
star, e.g.~\cite{ref36,PhysRevD.99.083016,Reyes:2018bee,asteroseismology}, GW polarization, e.g.~\cite{PhysRevD.100.064010,PhysRevD.91.082002}, black-hole ringdown, e.g.~\cite{PhysRevD.99.123029,PhysRevLett.123.111102},
application of machine learning techniques to GW data analysis, e.g.~\cite{PhysRevD.100.063015,PhysRevD.101.104003}, search for GW lensing effects, e.g.~\cite{Hannuksela_2019,lensing} and many other applications to
astrophysics and cosmology, e.g.~\cite{ref35,PhysRevD.100.104039,Finstad_2018,10.1093/mnras/staa1120,Chen_2020}. 
The list of projects goes beyond published scientific research and also includes student projects, academic courses, and art installations.\footnote{See \url{http://gw-openscience.org/projects/} for the list of scientific papers and projects.}

This paper is organized as follows. The section \ref{sec:methods}  provides insights about how the data are collected and calibrated, about data quality and simulated signal injections. 
The GWOSC file format and content are described in the section \ref{sec:record}, while the section \ref{sec:usage}  gives suggestions on the tools that can be used to guide the analysis of the GW data.

\section{Methods}
\label{sec:methods}
The Advanced LIGO~\cite{aasi15:_advan_ligo} and Advanced Virgo~\cite{acernese15:_advan_virgo} detectors are enhanced Michelson interferometers (see a simplified description of the experimental layout in Fig.~3
of~\cite{abbott16:_obser} and Fig.~3 of~\cite{acernese15:_advan_virgo}). Each detector has two orthogonal arms of equal length $L_{x} = L_{y} = L$, each with two mirrors acting as test masses and forming a Fabry--Perot optical
cavity. The arm length is $L=4$ km for LIGO, and $L=3$ km for Virgo. Advanced LIGO consists of two essentially identical detectors at Hanford, Washington and Livingston, Louisiana, while the Advanced Virgo
detector is located in Cascina near Pisa, Italy.

When GWs reach Earth, they alter the detector arm lengths, stretching or contracting each one according to the wave's direction, polarization and phase. This induces a time-dependent differential arm length change
$\Delta L = \delta L_{x}- \delta L_{y} = h L$, proportional to the GW strain amplitude $h$ projected onto the detector (see e.g.,~\cite{Maggiore} chap.~9, p.~470). Photodiodes continuously sense the differential length variations by measuring the
interference between the two laser beams that return to the beam splitter from the detector arms.

While Advanced LIGO and Advanced Virgo follow a similar general scheme, each facility has a specific, though closely related, design. We refer the reader to the following references for details about the technical
developments on the instrumentation and instrument controls that play a major part in reaching the sensitivities obtained during the O1 and O2 observing runs. For Advanced LIGO those include the light source (a
pre-stabilized laser)~\cite{Kwee:12,5866459420110301}, the main optics~\cite{Liu:13,Palashov:12,10.1063/1.3695405,Arain:08,Mueller:2015,0264-9381-24-2-008,Granata16,Pinard17}, the signal recycling mirror (used to optimize
the GW signal extraction)~\cite{PhysRevLett.116.131103,aasi15:_advan_ligo,PhysRevD.65.042001}, the optics suspension and seismic isolation
systems~\cite{0264-9381-31-10-105006,0264-9381-31-6-065010,Tokmakov20121699,0264-9381-29-12-124009,0264-9381-29-23-235004,0264-9381-29-11-115005,0264-9381-29-3-035003,10.1063/1.3532770,0264-9381-32-1-015004,Matichard2015273,Matichard2015287,0264-9381-32-18-185003,0264-9381-31-23-235001,0264-9381-21-9-003},
the sensing and control strategies~\cite{0264-9381-27-8-084026,Staley:15,PhysRevLett.114.161102}, the automation system~\cite{rsi.10.1063/1.4961665}, and various techniques for the mitigation of optical contamination,
stray light and thermal effects~\cite{10.1117/12.2047327,10.1142/9789814374552_0312,Brooks:09,0264-9381-19-7-377}.

For Advanced Virgo~\cite{acernese15:_advan_virgo,statusVirgo} a similar list includes the high reflective coatings of the core optics~\cite{coatingVirgo,coatingAmato}, the locking, control and thermal compensation
systems~\cite{lockVirgo,ControlVirgo,TCS}, and the mitigation of magnetic and seismic noises~\cite{MagNoiseVirgo,doi:10.1063/1.5045397,SeismVirgo,MultiSAS}.

When the detectors are taking data in their nominal configuration, they are said to be in \emph{observing mode} or \emph{science mode}. 
This condition does not occur all the time for various technical reasons. 
For example, the Fabry--Perot cavities included in the detector arms have to be kept at resonance together with the power and signal recycling cavities~\cite{Staley_2014}. 
There are periods when the control loops fail to maintain the instrument on this working point causing a non-observing period.  
Other possible reasons for non-observing include maintenance periods and environmental effects like earthquakes, wind and the microseismic ground motion arising from ocean storms~\cite{transLIGO,envLIGO}.

The time percentage during which the detectors are in science mode is called \emph{duty cycle} or \emph{duty factor}. 
During O1 the LIGO detectors had individual duty factors of 64.6\% for Hanford and 57.4\% for Livingston, while in O2 it was 65.3\% and 61.8\%, respectively. Virgo operated with a duty factor of 85.1\% during O2 (see table
1 of~\cite{observing_scenario}). 

If we define the \emph{network duty factor} by the time percentage during which all the detectors in the network are in science mode simultaneously,
we find 42.8\% for the LIGO network during O1 and 46.1\% during O2~\cite{LIGOScientific:2018mvr}. For the LIGO--Virgo network it was 35\%.

\begin{figure}[htb]
  \centering
  \includegraphics[width=0.8\linewidth]{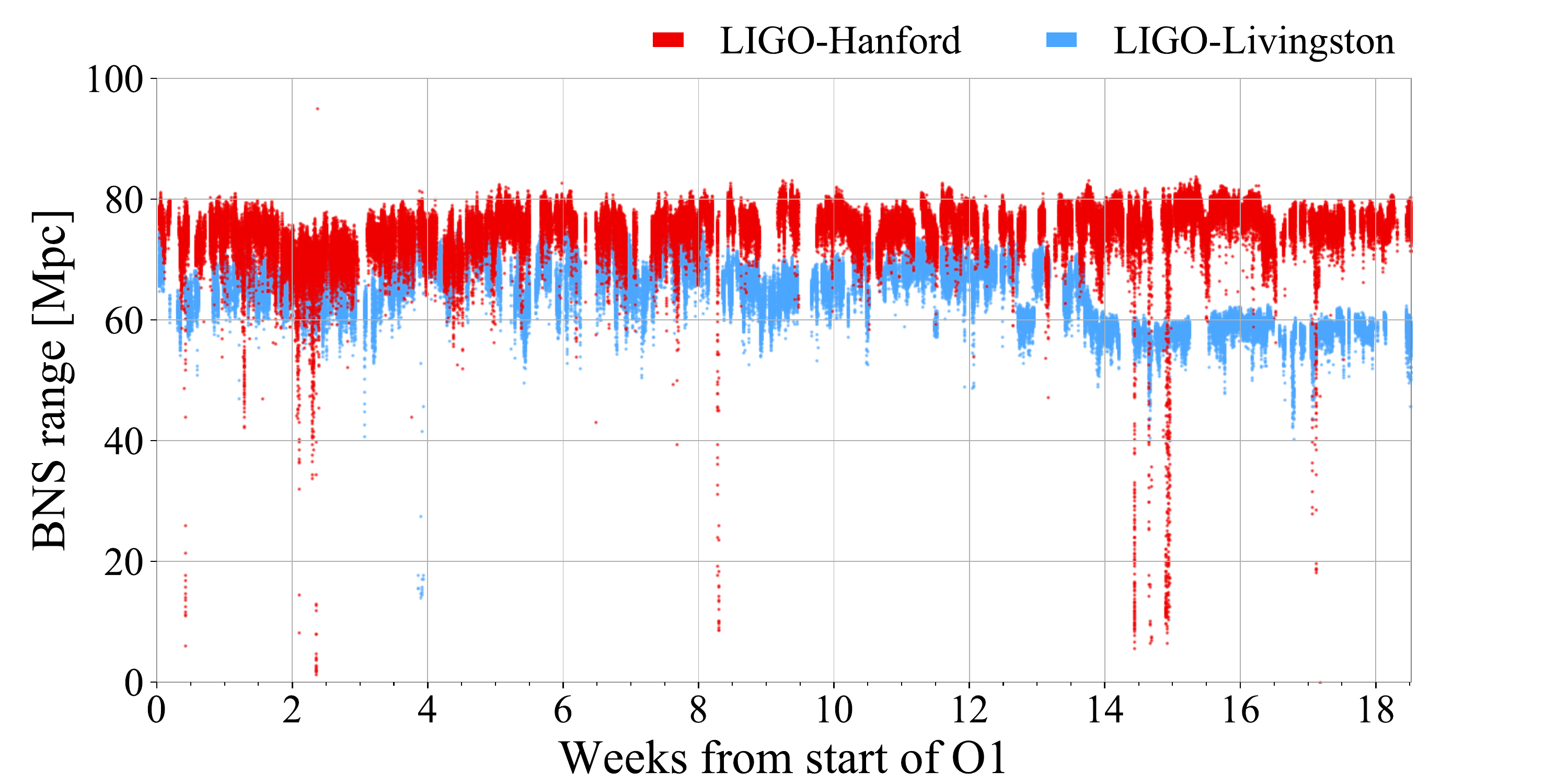}
  \includegraphics[width=0.8\linewidth]{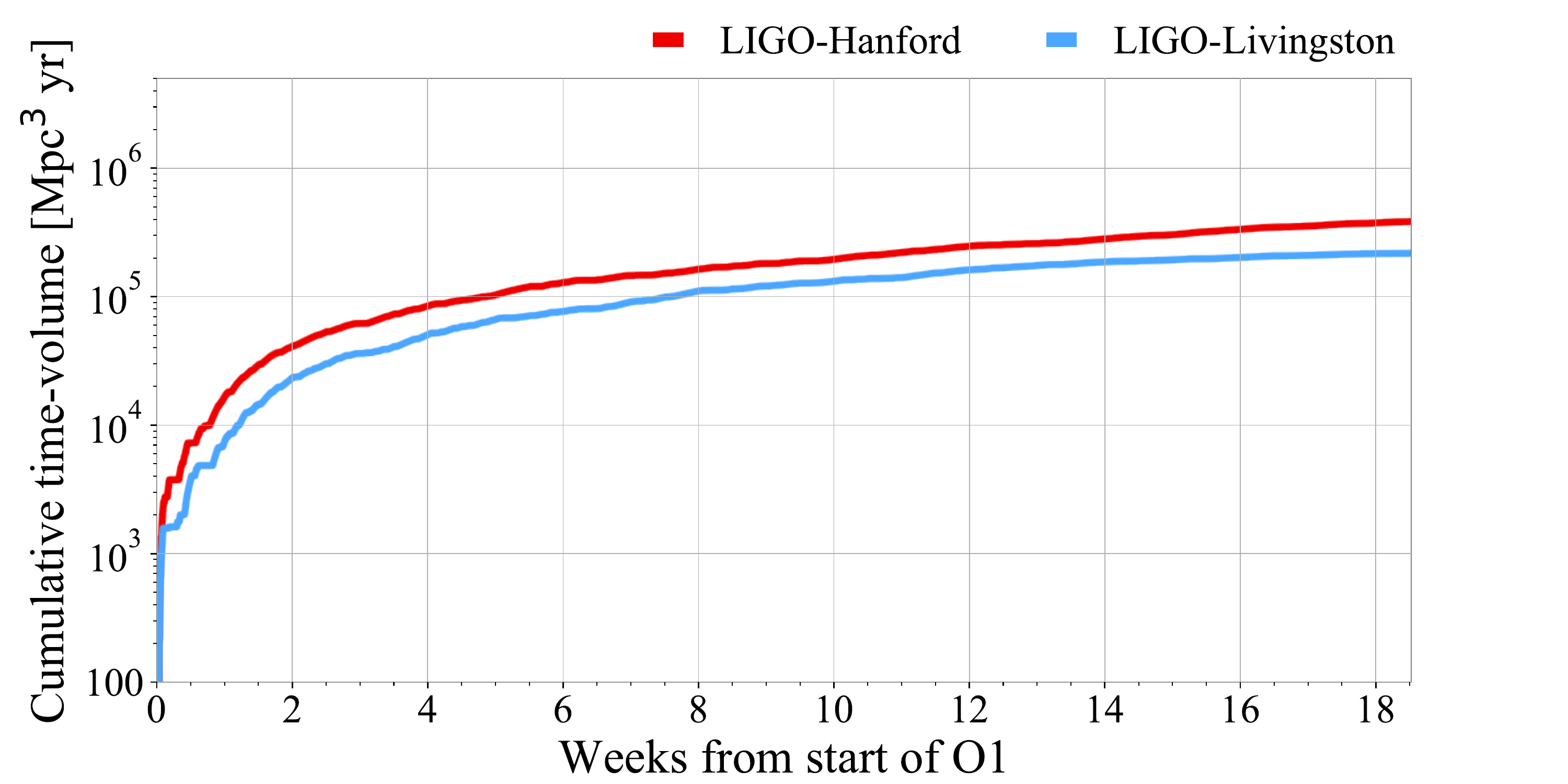} 
  \caption{Upper plot: O1 sensitivity of the Livingston and Hanford detectors to GWs as measured by the BNS range (in megaparsecs) to binary neutron-star mergers averaged over all sky positions and source
orientations~\cite{Finn:1992xs}. Lower plot: cumulative time--volume (assuming an Euclidean geometry appropriate for small redshifts) of the Livingston and Hanford detectors during O1, obtained by multiplying the observed
astrophysical volume by the amount of time spent observing.}
  \label{fig:range_vt_O1}
\end{figure}

\begin{figure}[htb]
  \centering
  \includegraphics[width=0.8\linewidth]{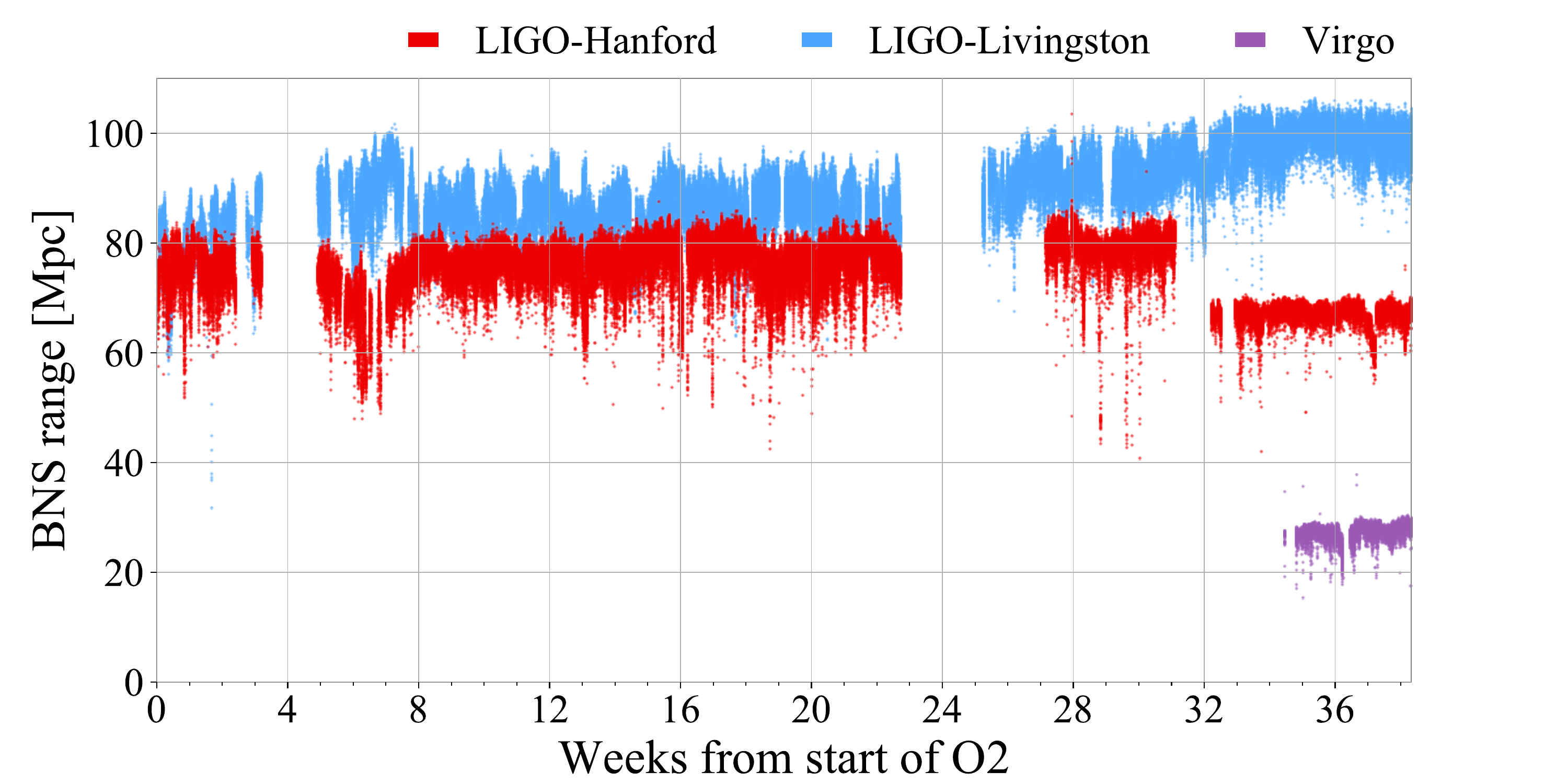}
  \includegraphics[width=0.8\linewidth]{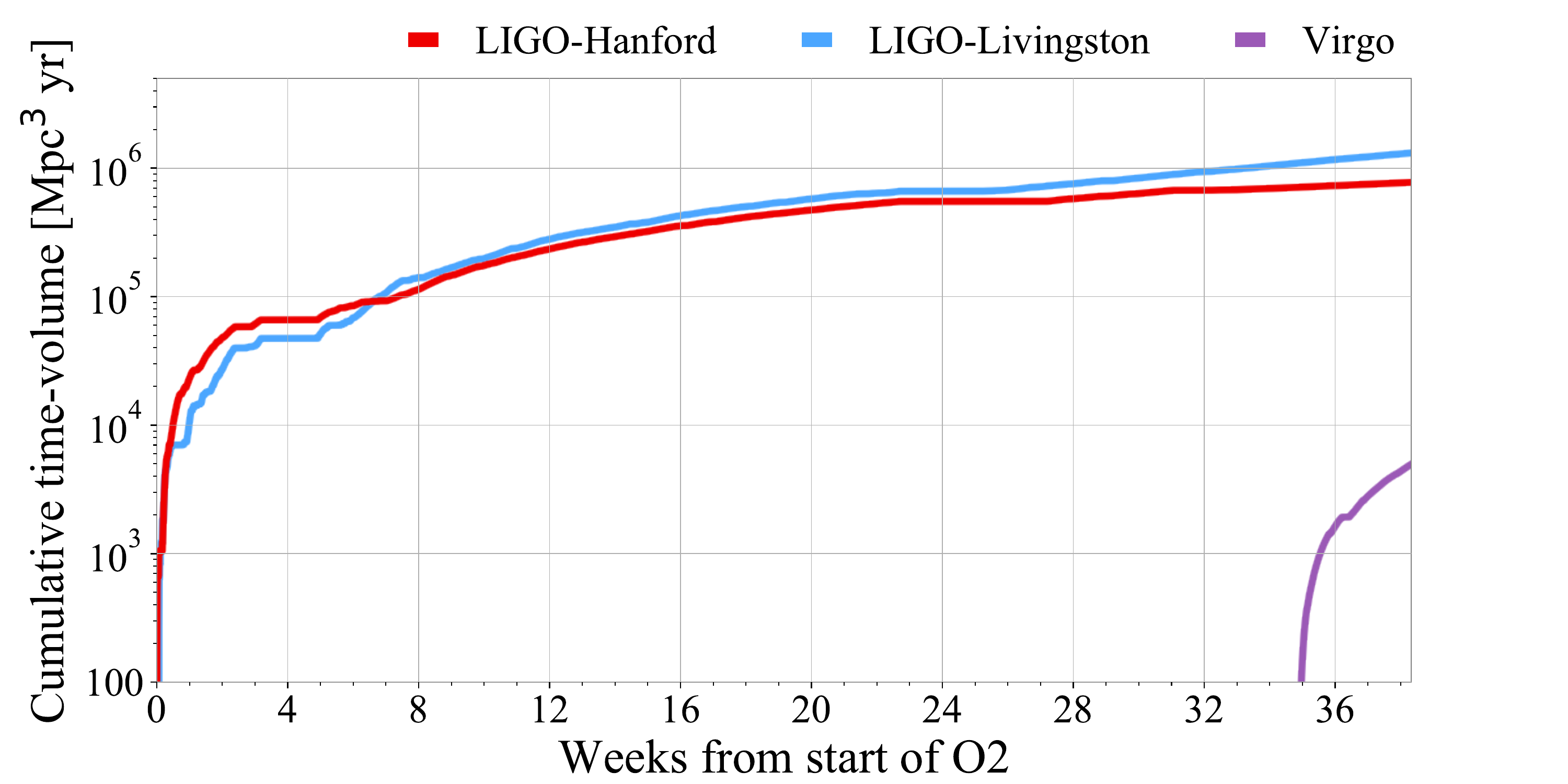} 
  \caption{Upper plot: O2 sensitivity of the Livingston, Hanford and Virgo detectors to GWs as measured by the BNS range (in megaparsecs) to binary neutron-star mergers averaged over all sky positions and source
orientations~\cite{Finn:1992xs}. Lower plot: cumulative time--volume (assuming Euclidean geometry appropriate for small redshifts) of the Livingston, Hanford and Virgo detectors during O2, obtained by multiplying the
observed astrophysical volume by the amount of time spent observing. Although Virgo has a lower BNS range and cumulative time--volume, its contribution is crucial for source localization and astrophysical parameter
estimation.}
  \label{fig:range_vt_O2}
\end{figure}

It is customary to quantify the detector sensitivity by the \textit{range}~\cite{Finn:1992xs,Chen:2017wpg}, i.e.,~the distance to which sources can be observed. 
In Figs. \ref{fig:range_vt_O1} and \ref{fig:range_vt_O2}, the \textit{BNS range} is calculated assuming that the observed source is a coalescence of compact objects of masses of 1.4 $\textup{M}_\odot$ each, the observation has a minimum
threshold in signal-to-noise ratio (SNR) of 8, and the range is averaged over all possible sky locations and orientations of the source, following~\cite{Finn:1992xs}. 
The figures contain also the equivalent cumulative time--volume~\cite{Chen:2017wpg} obtained by multiplying the amount of time spent observing by the observed astrophysical volume as defined by the range.
The sharp drops in the \textit{BNS range} are typically due to transient noise in the interferometer limiting its sensitivity, while the gaps are due to non-observing periods. In particular, during O2, there were two long breaks, one for end-of-year holidays and another to make improvements to the detectors. 
At the end of both runs there was a sensitivity drop in one of the LIGO detectors. For O1, a drop in sensitivity at LIGO Livingston was caused by electronics noise at one of the end stations while, for O2,  a drop in sensitivity at LIGO Hanford was due to electrostatic charging of the test mass optics caused by an earthquake in Montana.

The plots in  Figs. \ref{fig:range_vt_O1} and \ref{fig:range_vt_O2} are indicative of the performance of the individual detectors.\footnote{These figures are produced with data calibrated using the procedure described in the next section.} However, observations are performed jointly by Advanced LIGO and Advanced Virgo as a network. Roughly speaking, the sensitivity of the global network is determined by that of the second most sensitive detector operating at any time. 
Despite the lower BNS range and cumulative time--volume for Virgo, its contribution has been important for astrophysical parameter estimation, especially in determining source localization and
orientation~\cite{Abbott_2019}.
For instance, GW170814 and GW170817 were localized by the three-detector network within a few tens of square degrees while the other events were localized by the two-detector network in sky areas ranging from a few hundreds to several thousands of square degrees. 

Note that the sensitive distance depends strongly on the source mass, and can be much higher (up to gigaparsecs) for higher-mass BBH systems (see e.g.~Fig. 1 of Ref.~\cite{PhysRevD.100.064064}).


The next sections provide details on the calibration, the detector noise characterization, the data quality and signal injections.

\subsection{Calibration}

The differential arm length read-out of the interferometer is recorded digitally through a data acquisition system~\cite{aasi15:_advan_ligo,acernese15:_advan_virgo,0264-9381-27-8-084025}. The LIGO and Virgo data
acquisition systems acquire data at sampling rates $f_s = 16\,384~{\rm Hz}$ and $20\,000~{\rm Hz}$, respectively. The Virgo data is digitally converted to the same sampling rate as LIGO prior to any analysis. 

A calibration procedure~\cite{0264-9381-27-21-215001,0264-9381-27-8-084024,0264-9381-26-24-245011,Abadie2010223,VirgoCal1,VirgoCal2} is applied to produce the dimensionless strain from the differential arm length read-out. 
For both the Advanced LIGO and Advanced Virgo detectors, the calibration procedure creates a digital time series, $h(t)$, from the detector control system channels. 
Details of the production and characterization of $h(t)$ can be found in~\cite{2018CQGra..35i5015V,Acernese:2018bfl}. The calibration uncertainty estimation and residual systematic errors are discussed
in~\cite{Acernese:2018bfl,2017PhRvD..96j2001C,PhysRevD.95.062003}. 
The strain time series include both detector noise and any astrophysical signal that might be present.

Multiple versions of the calibrated data are produced as more precise measurements or instrument models become available.
A first strain $h(t)$ is initially produced online using calibration parameters measured just before the observing period starts.
This data stream is analyzed within a few seconds to generate alerts when an event is detected thus allowing follow-up observations by other facilities~\cite{alerts}.
Other offline versions of the calibration is produced later, offline, to include improvements to the calibration models or filters,
to resolve dropouts in the initial online version or after applying noise cleaning procedures.
This data stream is used in the production of the final search results, e.g.,~the final event catalog.

For the O1 and O2 science runs, we released the final offline version, that has the most precise uncertainties and after applying available noise cleaning procedures.
The calibration versions differ for the single event data releases depending on whether they pertain to the initial publication of the event (early version)
\cite{gwosc-GW150914,gwosc-LVT151012,gwosc-GW151226,gwosc-GW170104,gwosc-GW170608,gwosc-GW170814,gwosc-GW170817} or to the catalog GWTC-1 publication (final version)~\cite{gwosc-GWTC-1}.

The detector strain $h(t)$ is only calibrated between 10~Hz and 5000~Hz for Advanced LIGO~\cite{2017PhRvD..96j2001C,2018CQGra..35i5015V} and 10~Hz and 8000~Hz for Advanced Virgo~\cite{Acernese:2018bfl}. Any apparent
signal outside this range cannot be trusted because it is not a faithful representation of the GW strain at those frequencies. This part of the spectrum where the data are not calibrated corresponds to the regions where
the measurement noise increases rapidly, thus drastically reducing the chance for observing GWs.

\subsection{Detector noise characterization and data quality}

\begin{figure}[h]
  \centering
  \includegraphics[width=0.8\linewidth]{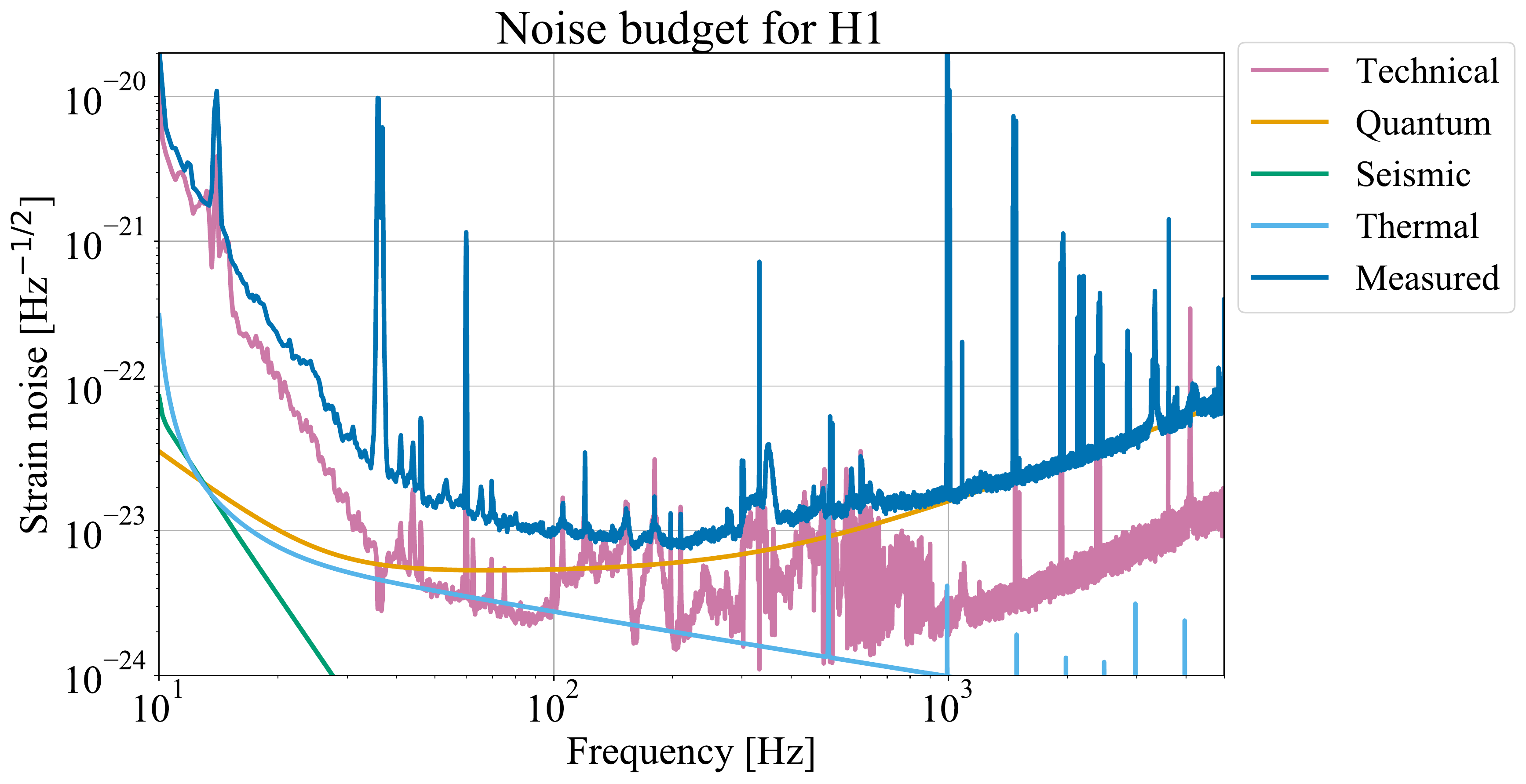}
  \includegraphics[width=0.8\linewidth]{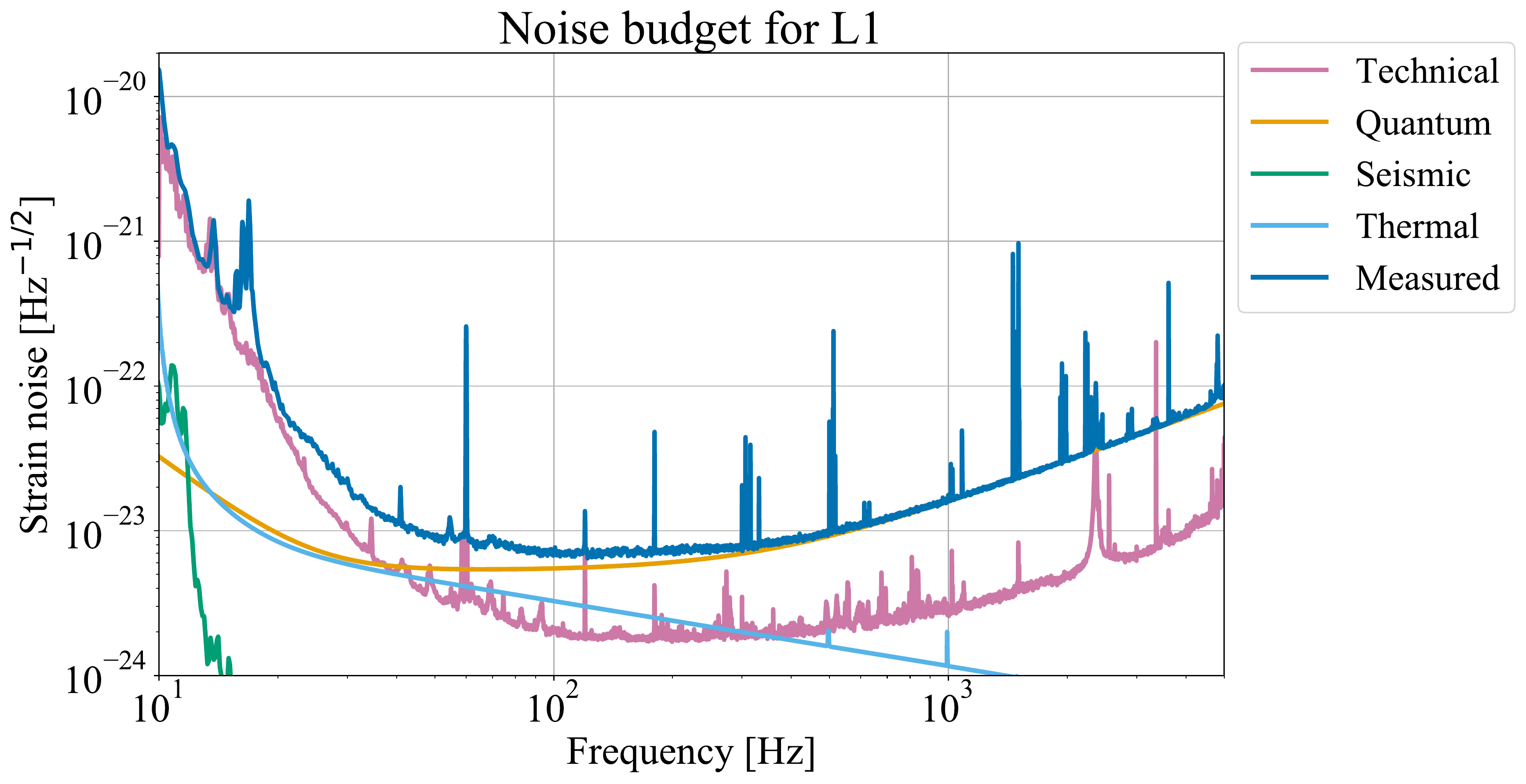} 
  \caption{Sensitivities of the Advanced LIGO detectors during the second observation run (O2), expressed as the equivalent strain noise spectrum of each detector (the blue ``Measured'' curves). 
Also shown are the known contributors to the detector noise, which sum to the measured spectrum across much, but not all of the frequency band 
(i.e.\, the measured noise spectrum is not fully explained by all known sources of noise).
The quantum noise includes both shot noise (dominant at higher frequencies) and radiation pressure noise (dominant at lower frequencies).
Thermal noise includes contributions from the suspensions, the substrate and coatings of the test masses.
Seismic noise is computed as the ground displacement attenuated through the seismic isolation system and the suspensions chain.
The seismic curves differ for H1 and L1 as actual seismic data were used for L1 while the H1 curve is a model that also includes Newtonian noise.
Technical noise includes angular and length sensing/control noise for degrees of freedom that are not related to the differential arm length measurement, and other sub-dominant noises such as laser frequency, intensity and beam jitter noise, sensor and actuation noise, and Rayleigh scattering by the residual gas.
The strong line features are due to the violin modes of the suspension wires, other resonance modes of the suspensions, the AC power line and its harmonics, and the calibration lines. Examples of similar plots for other
data taking runs can be found in~\cite{PhysRevLett.116.131103,Rana}. These noise spectra do not include any of the post-data collection noise subtraction mentioned in the text.
}
  \label{fig:noiseLIGO}
\end{figure}

\begin{figure}[h]
  \centering
  \includegraphics[width=0.9\linewidth]{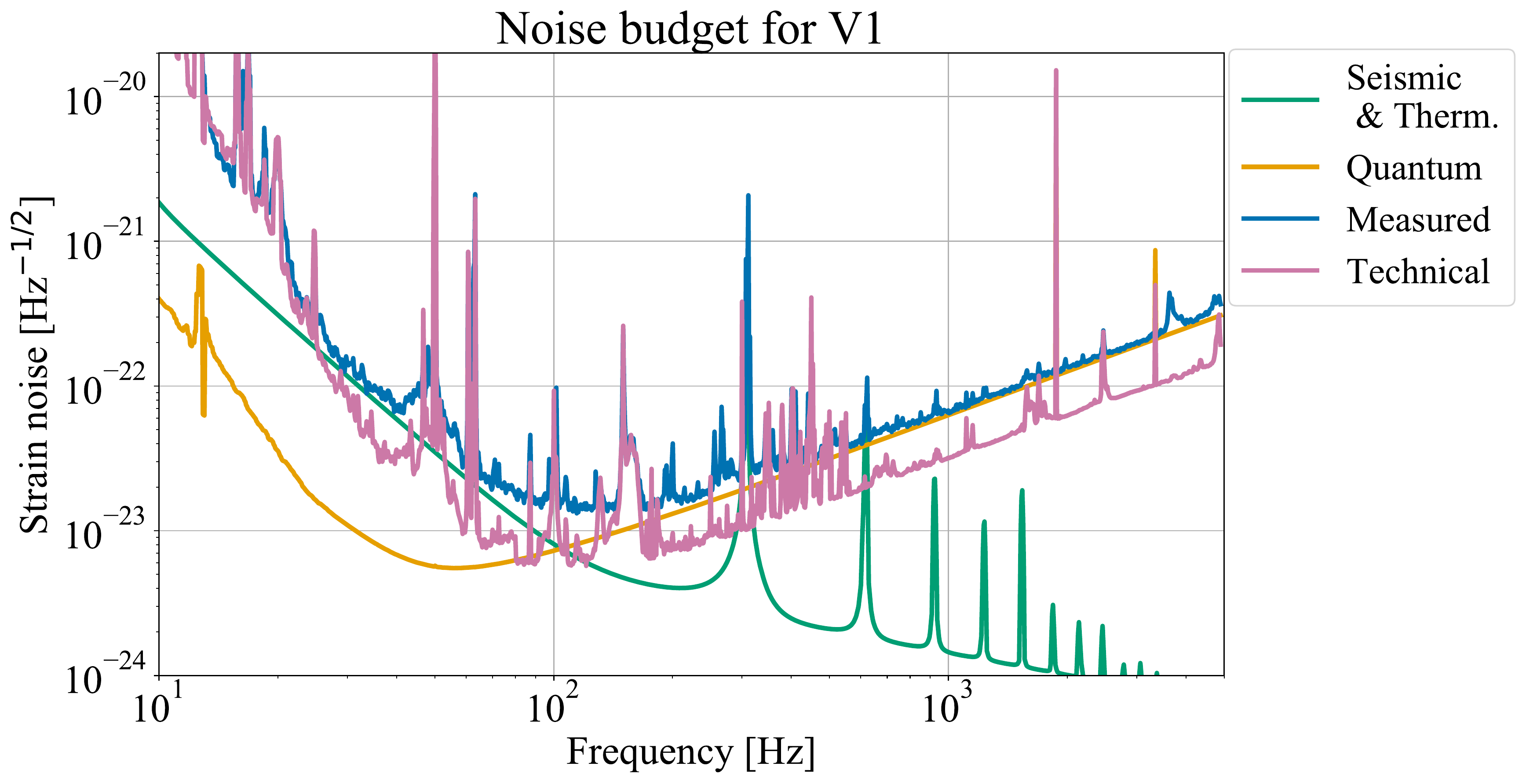}
  \caption{Sensitivity of the Advanced Virgo detector during the O2 observation run. The meaning of the noise source contributions is the same as in Fig.~\ref{fig:noiseLIGO}, except for the seismic and thermal noises that are combined in this case and for the Newtonian noise which is not included. These noise spectra do not include any of the post-data collection noise subtraction mentioned in the text.}
  \label{fig:noiseV1}
\end{figure}

The strain measurement is impacted by multiple noise sources, such as quantum sensing noise, seismic noise, suspension thermal noise, mirror coating thermal noise, and local gravity gradient noise produced by seismic waves
(called Newtonian noise)~\cite{aasi15:_advan_ligo}. The noise budget plot for Advanced LIGO during O1 can be found in~\cite{PhysRevLett.116.131103}.
In Figs.~\ref{fig:noiseLIGO} and \ref{fig:noiseV1} the noise budget for O2 is shown for Advanced LIGO and Advanced Virgo, respectively. 

The plots show the measured noise spectrum and the contribution from various known noise sources.\footnote{Other useful references for the detector sensitivity are~\cite{sensiO1} for O1 and~\cite{LIGOScientific:2018mvr} for
O2.} The noise spectra indicate that the dominant noises rise steeply at high and low frequencies. This opens an observational window between tens of Hz and a few kHz. Data analysis pipelines that are used to search for
gravitational-wave signals usually concentrate on frequency intervals smaller than the full calibrated bandwidth to avoid the high noise level at the extremes of this band.

The strain data are high-pass filtered at 8~Hz to avoid a number of digital signal processing problems related to spectral dynamic range and floating point precision limitation that may occur downstream when searching in the data.\footnote{See \url{https://www.gw-openscience.org/yellow_box/} and in particular the example showing the 8~Hz roll-off at \url{https://www.gw-openscience.org/static/images/ASDs_GW150914_1~Hz.png}.} 

As shown in Figs.~\ref{fig:noiseLIGO} and \ref{fig:noiseV1}, the data contain spectral lines that can complicate searches for signals in those frequency bands \cite{LIGOScientific:2019hgc}. These lines include calibration lines, power line harmonics, ``violin'' modes (resonant frequencies of mirror suspension fibers), other known instrumental lines, unknown lines and also evenly spaced combs of narrow lines, typically in exact multiples of some
fundamental frequency. Further details on spectral lines during O1 and O2 can be found in~\cite{lineLIGO,lineVirgo} as well as on the GWOSC web pages.\footnote{\url{http://gw-openscience.org/o1speclines} and
\url{http://gw-openscience.org/o2speclines}.}

The detector sites are equipped with a large number of sensors that monitor both the instrumental and environmental state (see~\cite{transLIGO,Aasi:2012wd,envLIGO} for details). The measurements performed by these sensors
are recorded in \emph{auxiliary channels} that are crucial for diagnosing instrument faults or for identifying environmental perturbations.
Non-Gaussian transient noise artifacts, called \emph{glitches}, can mask or mimic true astrophysical signals~\cite{transLIGO}. Auxiliary channels provide a useful source of information for the characterization of glitches,
and their mitigation. Glitches are caused by anomalous behavior in instrumental or environmental channels that couple into the GW channel. The observation of coincident glitches between the GW and auxiliary channels
provides a mechanism for rejecting a detected (potential) event in the former as not astrophysical in origin.
The large volume of auxiliary data (hundreds of thousands of auxiliary channels) are inspected (see~\cite{transLIGO,Aasi:2012wd,envLIGO} for details) and distilled into 
\emph{data quality vetoes} that allow identification of times that are unsuitable for analysis or are likely to produce false alarms. 
Veto conditions are determined using systematic studies to remove glitches with high efficiency and limited loss of observation time~\cite{transLIGO}.
As an example, vetoes discard glitches from electronics faults, photodiode saturations, analog-to-digital converter (ADC) and digital-to-analog converter (DAC) overflows, elevated seismic noise and computer failures. The
data quality vetoes are used by the GW searches to reduce the noise background~\cite{transLIGO} (see Sec. \ref{sec:DqInj} for search-related usage information).

Different categories of data quality are defined according to the severity level and degree of understanding of the noise artifact. Data flagged as invalid due to severe detector malfunctioning, calibration error, or data
acquisition problems, as described in~\cite{vetoO1} are typically not used for data analysis and are replaced by \texttt{NaN}s in the GWOSC data releases. We elaborate further on the various data quality categories and
their usage in the section \ref{sec:record}.\footnote{See also \url{http://gw-openscience.org/o1_details} and \url{http://gw-openscience.org/o2_details}.}

Auxiliary channels are also used to subtract post-facto some well identified instrumental noise from the GW strain data. A procedure based on a linear coupling model~\cite{O2clean} computes the transfer function that
couples the witness channels to $h(t)$ and subtracts the contributing noise from the strain amplitude.
This procedure was used during the second observing run in Advanced LIGO data. 
It achieved an increase of up to 30\%  of the detector sensitive volume to GWs for a broad range of compact binary systems and was most significant for the LIGO-Hanford detector~\cite{PhysRevD.98.084016}. In some cases
data are available both before and after noise subtraction is applied (for example in the case of GW170817~\cite{gwosc-GW170817}). 
\subsection{Signal injections}

In addition to data quality, some metadata provide information about \textit{hardware injections}~\cite{inj} inserted into the detector data for testing and calibration. The detectors' test masses (interferometer mirrors)
are physically displaced by an actuator in order to simulate the effects of a GW.\footnote{Calibration lines mentioned earlier are generated using the same process.} A simulated GW signal is introduced into the detector
control system yielding a response which mimics that of a true GW. The analysis of a data segment that includes an injection allows an end-to-end test of the ability for the analysis procedure to detect and characterize
the GW strain signal.

Hardware injections are also used for detector characterization to check that the auxiliary channels used for vetoes do not respond to gravitational-wave-like signals. This is a \textit{safety} check since a channel that has no sensitivity to GWs is considered safe for use when constructing a veto.
It is clearly important to keep a record of injections to avoid any confusion with real events. In the section \ref{sec:record}  we describe how this bookkeeping is done.\footnote{See the GWOSC web page \url{http://gw-openscience.org/o1_inj} and \url{http://gw-openscience.org/o2_inj}.}

\section{Data records}
\label{sec:record}

GW open data are distributed under the Creative Commons Attribution International Public License 4.0\footnote{\url{https://creativecommons.org/licenses/by/4.0/legalcode}} through the GWOSC web pages.\footnote{\url{http://gw-openscience.org/data/}}
The files can be directly downloaded one by one from this web page. However, to download large amounts of data (such as an entire observing run) the use of the distributed filesystem
\texttt{CernVM-FS}~\cite{CVMFS,CVMFSpaper} is preferred.\footnote{For installation instructions, see \url{http://gw-openscience.org/cvmfs/}} Once installed, this filesystem allows access to GWOSC data as files in a
directory tree mounted locally on the user's computer.

The calibrated strain data of O1~\cite{gwosc-O1} and O2~\cite{gwosc-O2} described in this paper are conveniently divided into files of 4096 s.  Short segments of 32~s and 4096~s duration for each GW event are also
released.\footnote{\url{http://gw-openscience.org/eventapi/}} The description of the data records that follows is valid both for single event release and for bulk data release.

The strain data are repackaged and resampled by GWOSC to make it more accessible to users both within the LVC and outside. Along with the native 16\,384~Hz sampling rate, the data on GWOSC are also made available at 4096
Hz.\footnote{In the rest of the paper the sampling rates will be indicated in kHz and rounded to the closest integer, i.e.\ 4 and 16~kHz means 4096 and 16\,384~Hz, respectively} The down-sampling is performed using the
standard decimation technique implemented in \texttt{scipy.signal.decimate}\footnote{This method applies an anti-aliasing filter based on an order-8 Chebyshev type I infinite impulse response (IIR) filter~\cite{IIR} before
decimation.} from the Python package \texttt{scipy}~\cite{scipy}. 
From the Nyquist--Shannon sampling theorem~\cite{Nyquist1,Nyquist2,shannon}, the largest accessible frequency is the Nyquist frequency equal to half of the sampling rate $f_s$. This should be kept in mind when
choosing the sampling rate to download from GWOSC, and in general when analyzing these files; in particular, because of the anti-aliasing filter's roll-off, the data sampled at 4~kHz are valid only up to frequencies of
about 1700~Hz. 

The publicly released data are generated from data streams in the LIGO and Virgo data archives uniquely identified by a channel name and a frame type (an internal label that specifies the content of the files). 
For completeness, we give the provenance of the GWOSC data in Table \ref{tab:source} and list the channel names and frame types used to generate the O1 and O2 dataset discussed in this article. In this table and in the following, H1 and L1 indicate the two LIGO detectors (Hanford and Livingston respectively) while V1 refers to Virgo. 
Downsampling (for the 4~kHz dataset) and replacement with \texttt{NaN}s of bad quality or absent data are the only modifications of the original data.

\begin{table}[ht]
\centering
\caption{The channel names and frame types listed in this table are unique identifiers in the LIGO and Virgo data archives that allow tracing the provenance of the strain data released on GWOSC. The attribute \texttt{CLEAN} in H1 and L1 for O2 indicates that the noise subtraction procedure mentioned previously and described in \cite{O2clean} was used. The attributes \texttt{C02}, \texttt{DCS},  \texttt{DCH} and \texttt{Repro2A} refer to the calibration version.}
\vspace{10pt}
\label{tab:source}
\resizebox{\textwidth}{!}{\begin{tabular}{llll}
\hline
Run	&	Det.	&	Channel name	&	Frame type\\
\hline
\hline
O1	&	H1	&	\texttt{H1:DCS-CALIB\_STRAIN\_C02}			&	\texttt{H1\_HOFT\_C02}\\
O1	&	L1	&	\texttt{L1:DCS-CALIB\_STRAIN\_C02}			&	\texttt{L1\_HOFT\_C02}\\
\hline
O2	&	H1	&	\texttt{H1:DCH-CLEAN\_STRAIN\_C02}		&	\texttt{H1\_CLEANED\_HOFT\_C02}\\
O2	&	L1	&	\texttt{L1:DCH-CLEAN\_STRAIN\_C02}		&	\texttt{L1\_CLEANED\_HOFT\_C02}\\
O2	&	V1	&	\texttt{V1:Hrec\_hoft\_V1O2Repro2A\_16384Hz}	&	\texttt{V1O2Repro2A}\\
\hline
\end{tabular}
}
\end{table}  
\subsection{GWOSC file formats}

The GW open data are delivered in two different file formats:  \texttt{hdf} and  \texttt{gwf}. 
The Hierarchical Data Format \texttt{hdf}~\cite{hdf5} is a portable data format readable by many programming languages. 
The Frame format \texttt{gwf}~\cite{frame} is used internally by the GW community. 
In addition, the data associated with GW events are also released as plain text files containing two columns with the GPS time and the corresponding strain values. 

There are some differences in the structure of the file names between O1 and O2 due to the evolution of GWOSC itself. For O1 the name of the files has the structure: \textit{obs}-\textit{ifo}\_\texttt{LOSC}\_\textit{s}\_\texttt{V}\textit{n}-\textit{GPSstart}-\textit{duration}.\textit{extension}, where \textit{obs} is the observatory, i.e.~the site, so can have values L or H; \textit{ifo} is the interferometer and can have values H1 or L1; \texttt{LOSC} is the previous name of GWOSC, (the L in LOSC stands for LIGO); \textit{s} is the sampling rate in kHz with possible values 4 or 16; \textit{n} is the version number of the file (until now we have only one version, so only V1); \textit{GPSstart} is the starting time in GPS of the data contained in the file; \textit{duration} is the duration in seconds of this segment of data, which value is always 4096 in this case; the \textit{extension} can be gwf or hdf. The file names in O2 are instead of the type \textit{obs}-\textit{ifo}\_\texttt{GWOSC}\_\textit{ObservationRun}\_\textit{s}\texttt{KHZ}\_\texttt{R}\textit{n}-\textit{GPSstart}-\textit{duration}.\textit{extension}, with the same meaning of the italic letters, but in this case \textit{obs} and \textit{ifo} can have also the values V and V1, respectively, for Virgo data and we added the run name in the file names, so in this case \textit{ObservationRun} is O2.

The folders (or groups) included in the \texttt{hdf} files are:
\begin{itemize}
\item \emph{meta}: metadata of the file containing the following fields: 
\begin{itemize}
  \item \emph{Description}, e.g.~``Strain data time series from LIGO'',
  \item \emph{DescriptionURL}: URL of the GWOSC website,
  \item \emph{Detector}, e.g.~L1, and \emph{Observatory}, e.g.~L,\footnote{The observatory refer to the site and it is indicated by one letter, like L for Livingston. The addition of a number after the letter to indicate the detector, e.g.~L1, could be useful if multiple detectors are installed in the same site, as it was at the beginning of LIGO.}
  \item \emph{Duration}, \emph{GPSstart}, \emph{UTCstart}: duration and starting time (in GPS and UTC, respectively) of the segment of data contained in the file.
  
\end{itemize}
 
 In the O2 files it was decided to add also the \emph{StrainChannel} and \emph{FrameType} of the original files internally used by the LVC (i.e.~the content of Table \ref{tab:source}).
 
\item \emph{strain}: array of $h(t)$, sampled at 4 or 16~kHz depending on the file. For the times when the detector is not in science mode or the data does not meet the minimum required data quality conditions (see next section), the strain values are set to \texttt{NaN}s. The strain $h(t)$ is a function of time, so it is accompanied by the attributes \emph{Xstart} and \emph{Xspacing} defining the starting GPS time of the data contained in the array and the corresponding distance in time between the points of the array.

\item \emph{quality}: this folder contains two sub-folders, one for data quality and the other for injections, each including a bitmask to indicate at each second the status of the data quality or the injections and the description of each bit of the mask, i.e.~the content of~Tables \ref{tab:dq} and \ref{tab:inj} (see section \ref{sec:DqInj} for details).

\end{itemize}

The \texttt{gwf} files have a similar content but with a different structure. They contain 3 channels, one for the strain data, one for the data quality and one for the injections. The channel names differ slightly in O1 and O2 as described in Table \ref{tab:channels}. Note that the original files produced internally, whose channel names are listed in Table \ref{tab:source}, contain only the strain channel, while the GWOSC files include also the data quality and injection information in the same file. 

\begin{table}[ht]
\centering
\caption{Channel names of the GWOSC frame (\texttt{gwf}) files. In the name, \textit{ifo} is a place holder for the interferometer name, i.e. H1, L1 or V1, and \textit{s} the sampling rate in kHz. The \texttt{R1} substring represents the revision number of the channel name so it will become \texttt{R2} in case there is a second (revised) release, and so on.}
\vspace{10pt}
\label{tab:channels}
\begin{tabular}{lll}
\hline
				&	O1 (4 kHz sampling)				&	O1 (16 kHz sampling) and O2\\
\hline
\hline
Strain			&	\textit{ifo}\texttt{:LOSC-STRAIN}	&	\textit{ifo}\texttt{:GWOSC-}\textit{s}\texttt{KHZ\_R1\_STRAIN}	\\
\hline
Data quality mask 	&	\textit{ifo}\texttt{:LOSC-DQMASK}	&	\textit{ifo}\texttt{:GWOSC-}\textit{s}\texttt{KHZ\_R1\_DQMASK}	\\
\hline
Injections mask		&	\textit{ifo}\texttt{:LOSC-INJMASK}	&	\textit{ifo}\texttt{:GWOSC-}\textit{s}\texttt{KHZ\_R1\_INJMASK}	\\
\hline
\end{tabular}
\end{table}  
\subsection{Data quality and injections in GWOSC files}
\label{sec:DqInj}

Several types of searches are performed on the LIGO and Virgo data. Those searches are divided into four families named after the type of signals they target: Compact binary coalescences (\texttt{CBC}), GW bursts (\texttt{BURST}), continuous waves (\texttt{CW}) and stochastic backgrounds (\texttt{STOCH}).

\texttt{CBC} analyses (see e.g.,~\cite{pycbc-github,Usman:2015kfa,Sachdev:2019vvd,Messick:2016aqy,LIGOScientific:2018mvr,GW150914,PhysRevLett.123.161102,PhysRevD.100.064064}) seek signals from merging neutron stars and
black holes by filtering the data with waveform templates.
\texttt{BURST} analyses (see e.g.,~\cite{Klimenko:2015ypf,PhysRevD.100.024017,CCSN,PhysRevD.95.104046,PhysRevD.94.044050}) search for generic GW transients with minimal assumption on the source or signal morphology by
identifying excess power in the time--frequency representation of the GW strain data.
\texttt{CW} searches (see e.g.,~\cite{CW,PhysRevD.99.122002,CW1,CW2}) look for long-duration, continuous, periodic GW signals from asymmetries of rapidly spinning neutron stars. 
\texttt{STOCH} searches (see e.g.,~\cite{PhysRevLett.118.121101,PhysRevD.100.061101}) target the stochastic GW background signal which is formed by the superposition of a wide variety of independent and unresolved sources
from different stages of the evolution of the Universe.

Due to the fundamental differences among these searches, some types of noise are problematic only for one or two types of search. 
For this reason, the data quality related to transient noises depends on the search type. It is provided inside the GWOSC files for the two GW transient searches \texttt{CBC} and \texttt{BURST}, that are most sensitive to
this type of noise. The data quality information most relevant for \texttt{CW} and \texttt{STOCH} searches is in the frequency domain and it is provided as lists of instrumental lines in separate
files~\cite{H1linesO1,L1linesO1,H1linesO2,L1linesO2,V1linesO2}.

Data quality and signal injection information for a given GPS second is indicated by bitmasks with a 1-Hz sampling rate. The bit meanings are given in Tables \ref{tab:dq} and \ref{tab:inj} for the data quality and injections, respectively.
To describe data quality, different \emph{categories} are defined. 
For each category, the corresponding bit in the bitmask shown in Table \ref{tab:dq} has value 1 (good data) if in that second of time the requirements of the category are fulfilled, otherwise 0 (bad data). 

The meaning of each category is the following:
\begin{description}
\item[\texttt{DATA}] Failing this level indicates that LIGO and Virgo data are not available in GWOSC data because the instruments were not operating in nominal conditions. For O1 and O2, this is equivalent to failing Category 1 criteria, defined below. 
For these seconds of bad or absent data, \texttt{NaN}s have been inserted.
\item[\texttt{CAT1}](Category 1) Failing a data quality check at this category indicates a critical issue with a key detector component not operating in its nominal configuration. Since these times indicate a major known problem they are identical for each data analysis group. 
However, while \texttt{CBC\_CAT1} and \texttt{BURST\_CAT1} flag the same data, they exist separately in the dataset.
GWOSC data during times that fail CAT1 criteria are replaced by \texttt{NaN} values in the strain time series. The time lost due to these critical quality issues (\emph{dead time}) is: 1.683\% (H1) and 1.039\% (L1) of the run during O1; and 0.001\% (H1), 0.003\% (L1) and 0.053\% (V1) of the run during O2 (all the percentages have been calculated with respect to the periods of science mode).
\item[\texttt{CAT2}](Category 2) Failing a data quality check at this category indicates times when there is a known, understood physical coupling between a sensor/auxiliary channel that monitors excess noise, and the
strain channel~\cite{catO1}. The dead times corresponding to this veto for the \texttt{CBC} analysis are: 0.890\% (H1) and 0.007\% (L1) of the run during O1; 0.157\% (H1) and 0.090\% (L1) of the run during O2.  
The dead times corresponding to this veto for the \texttt{BURST} analysis are: 0.624\% (H1) and 0.021\% (L1) of the run during O1; 0.212\% (H1) and 0.151\% (L1) of the run during O2. \texttt{CAT2} was not used for Virgo in O2.
\item[\texttt{CAT3}] (Category 3) Failing a data quality check at this category indicates times when there is statistical coupling between a sensor/auxiliary channel and the strain channel which is not fully understood. This category was not used in O1 and O2 LVC searches, but it is still in the file format for historical reasons.
\end{description}

As an example,~\cite{catO1} gives the list of all sensors/auxiliary channels used to define the \texttt{CAT1} and \texttt{CAT2} flags for \texttt{BURST} and \texttt{CBC} around the event GW150914.

Data quality categories are cascading: a time which fails a given category automatically fails all higher categories. Since \texttt{CAT3} is not used in this specific case and only data passing \texttt{CAT1} are provided, there is only the possibility that the data pass or not \texttt{CAT2}.
However, the different analysis groups qualify the data independently: failing \texttt{BURST\_CAT2} does not necessarily imply failing \texttt{CBC\_CAT2}.

\begin{table}[ht]
\centering
\caption{Data quality bitmasks description. Data that are {\it not} present are replaced by \texttt{NaN} values in the strain time series.
\texttt{CBC\_CAT1} and \texttt{BURST\_CAT1} are equivalent (see the definition of \texttt{CAT1} in the text).}
\vspace{10pt}
\label{tab:dq}
\begin{tabular}{lll}
\hline
Bit	&	Short name	&	Description	\\
\hline
\hline
0	& 	\texttt{DATA}			&	Data present	\\
1 	&	\texttt{CBC\_CAT1} 		&	Pass \texttt{CAT1} test\\
2 	&	\texttt{CBC\_CAT2} 		&	Pass \texttt{CAT1} and \texttt{CAT2} test for \texttt{CBC} searches\\
3 	&	\texttt{CBC\_CAT3} 		&	Pass \texttt{CAT1} and \texttt{CAT2} and  \texttt{CAT3} test for \texttt{CBC} searches\\
4 	&	\texttt{BURST\_CAT1} 	&	Pass \texttt{CAT1} test\\
5 	&	\texttt{BURST\_CAT2} 	&	Pass \texttt{CAT1} and \texttt{CAT2} test for \texttt{BURST} searches\\
6 	&	\texttt{BURST\_CAT3} 	&	Pass \texttt{CAT1} and \texttt{CAT2} and  \texttt{CAT3} test for \texttt{BURST} searches\\
\hline
\end{tabular}
\end{table}  

The injection bitmask marks the injection-free times. 
Five different types of injections are usually performed: injections simulating signals searched for by \texttt{CBC}, \texttt{BURST}, \texttt{CW} and \texttt{STOCH} LVC pipelines, and injections used for detector characterization labeled \texttt{DETCHAR}.
For each injection type, the bit of the bitmask, whose meaning is described in Table~\ref{tab:inj}, has value 1 if the injection is not present, otherwise 0.

Virgo did not perform hardware injections during O2, therefore all the bits of the injection bitmask have value 1.

\begin{table}[ht]
\centering
\caption{Meaning of the injection bits}
\vspace{10pt}
\label{tab:inj}
\begin{tabular}{lll}
\hline
Bit	&	Short name				&	Description	\\
\hline
\hline
0	& 	\texttt{NO\_CBC\_HW\_INJ}			&	No \texttt{CBC} injections	\\
1 	&	\texttt{NO\_BURST\_HW\_INJ} 		&	No burst injections\\
2 	&	\texttt{NO\_DETCHAR\_HW\_INJ} 	&	No detector characterization injections\\
3 	&	\texttt{NO\_CW\_HW\_INJ}  		&	No continuous wave injections\\
4 	&	\texttt{NO\_STOCH\_HW\_INJ} 		&	No stochastic injections\\

\hline
\end{tabular}
\end{table}  

\section{Technical Validation}

The data repackaged for public use are validated by another independent internal team. In particular, this review team checks that:
\begin{itemize}
\item the strain vectors in the GWOSC \texttt{hdf} and \texttt{gwf} files are identical to machine precision to the corresponding strain vectors of the LVC main archives;
\item the data quality and injection information given to the user correspond to what is included in the original LVC data quality database. The user can get this information in two ways: the bitmask included in the GWOSC files and the \textit{Timeline} tool described in detail in the section \ref{sec:usage}. The output of both methods is checked against the database;
\item the documentation web pages and the content of the present article contain correct and comprehensive information.
\end{itemize}

The data files, the \textit{Timeline} and the web pages are released to the public once all those checks have been passed.

\section{Usage notes}
\label{sec:usage}
GW detectors are complex instruments, and their data reflect this complexity. 
For this reason, caution should be taken when searching for GW signals in the detector strain data, taking into account all the details about the usable frequency range, noise artifacts, data quality and injections
discussed in this paper and in the references. In particular, the application of all data quality flags described in the previous section does \textit{not} imply that the remaining data are free of transient noise
artifacts. The user can find guidance to analyze the GW data in the tutorials and tools collected in the GWOSC website and discussed in the next sub-sections. The data analysis techniques used to detect GW signals and
infer the source properties are described in~\cite{LIGOScientific:2019hgc} where good practices and advices to avoid common errors are also provided. The GWOSC website also contains basic information about the geographical
position\footnote{\url{http://gw-openscience.org/static/param/position.txt}} and the current status\footnote{\url{http://gw-openscience.org/detector_status/}} of the detectors.

\subsection{Timeline}

\begin{figure}[h]
  \includegraphics[width=\linewidth]{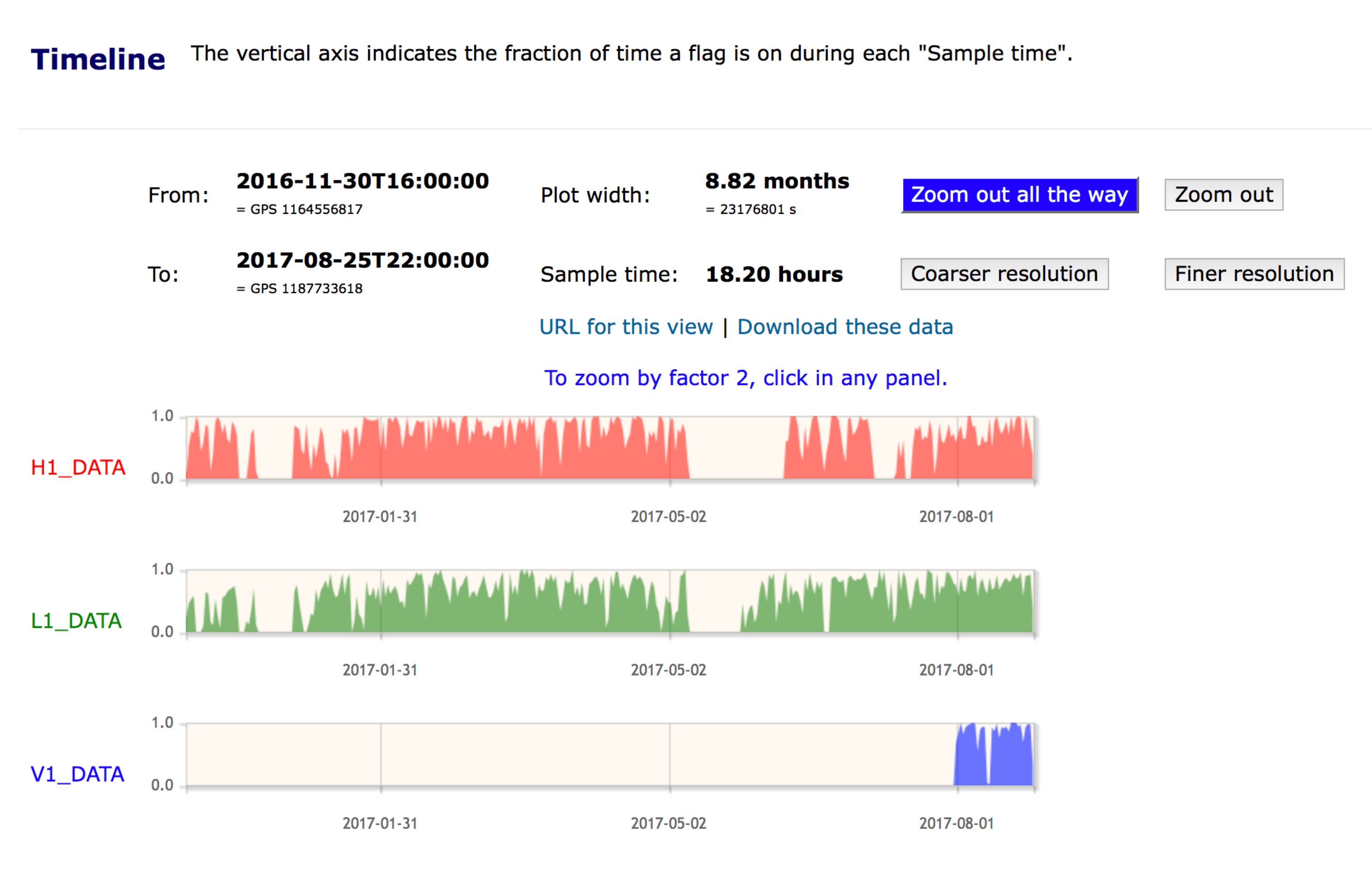}
  \caption{The GWOSC offers immediate access to duty cycle information for data quality and injection bits through the \textit{Timeline} (\protect\url{http://gw-openscience.org/timeline/}). By default, the time resolution is chosen to display the entire dataset. From there, one can zoom in to smaller timescales by clicking on the display.}
  \label{fig:timeline_O2}
\end{figure}

The LIGO and Virgo detectors are not always in observing mode and, even when they are, it is possible that data quality does not meet the requirements of a given analysis. 
For these reasons it is necessary to restrict analysis to valid \textit{segments} of data characterized by data quality information that indicates the data is acceptable for the desired analysis. 
 \textit{Timeline}\footnote{\url{http://gw-openscience.org/timeline/}} is a tool to provide a visual representation of available valid data segments over a time interval, together with the related information about data quality and presence of injected signals (see Fig. \ref{fig:timeline_O2} for an example with the O2 dataset). If the requested interval is short enough, this is shown at the time scale of seconds. 
For longer intervals, \textit{Timeline} shows the average value of the selected data-quality bit over nonoverlapping 2$^{n}$-second subintervals. 

Besides the visual representation, this tool allows the user to download the list of start and stop of the segments for a specific data quality category or injection type, and also the corresponding data.

\subsection{Courses, software packages and tutorials for GW data analysis}

On-line courses that provide an introduction to GW data analysis ranging from the basics to more advanced topics with hands-on exercises are available from the GWOSC website.\footnote{\url{http://gw-openscience.org/workshops/}} 
Those courses have been recorded during the GW Open Data Workshops. They include lectures on various aspects of GW science and are supported by many tutorials that can be used to understand how to read and analyze the data. 
The tutorials on the GOWSC website\footnote{\url{http://gw-openscience.org/tutorials/}} are in the form of \texttt{Jupyter} notebooks~\cite{Kluyver:2016aa}. They explain how to access the data, produce time--frequency
spectrograms, carry out matched-filtering searches, infer astrophysical parameters, and manipulate GW localization information. 
A few tutorials start from first principles and use generic and broadly used analysis software such as \texttt{scipy}~\cite{scipy}, but most are based on the specialized software packages and libraries that the LVC
developed to produce observational results and other scientific products.

A list of those packages is available on the GWOSC website\footnote{\url{http://gw-openscience.org/software/}} and includes:
\begin{itemize}
\item the light-weight application \texttt{readligo} to access data;
\item general purpose application software, such as the LSC Algorithm Library Suite (\texttt{LALSuite})~\cite{lalsuite} and the Python package \texttt{gwpy}~\cite{gwpy};
\item search-oriented software such as \texttt{pycbc}~\cite{pycbc-github,Usman:2015kfa}, \texttt{GstLAL}~\cite{gstlal} and Coherent Waveburst (\texttt{cWB})~\cite{Klimenko:2015ypf};
\item post-processing software for e.g.,~parameter estimation such as \texttt{bilby}~\cite{bilby}, \texttt{LALInference}~\cite{PhysRevD.91.042003} and Bayeswave~\cite{2015CQGra..32m5012C,PhysRevD.91.084034}.
\end{itemize}

All these packages are open source and freely distributed.

\subsection{Summary and additional information}

The LVC is committed to providing strain data from the LIGO and Virgo detectors to the public, 
according to the schedule outlined in the LIGO Data Management Plan~\cite{LIGO-DMP},
via the Gravitational Wave Open Science Center (GWOSC)~\cite{gwosc}.
They are also committed to providing a broad range of data analysis products to facilitate reproducing the results 
presented in their observational papers. Many of these data products are available through the LIGO Document Control Center (DCC); 
for example, data products associated with the GWTC-1 event catalog~\cite{LIGOScientific:2018mvr} can be found in
\cite{gwosc-GWTC-1} and~\cite{DCC-GWTC-1}. 
Many more data offerings are planned for the future. This includes the catalog of observed events and the bulk strain data from the 
LIGO/Virgo O3 run. More GWOSC Open Data Workshops are also planned.

All users of these data are welcome to sign up with the GWOSC User's Group at \url{https://www.gw-openscience.org/join/}.
Anyone who uses these data in publications and other public data products are requested to acknowledge GWOSC
by following the guidance in~\cite{gwosc-ack}.
Publications that acknowledge GWOSC will be listed in \url{https://www.gw-openscience.org/projects/}; email \texttt{gwosc@igwn.org} to make sure your publication(s) are included.

The Collaborations, and the GWOSC team, welcome comments and suggestions for improving these data releases and products,
and their presentation on the GWOSC website~\cite{gwosc}, via email to \texttt{gwosc@igwn.org}.
Questions about the use of these data products may also be sent to that email, and will be entered into our help ticket system.
More general questions about LIGO, Virgo, and GW science should go to \texttt{questions@ligo.org}. 

\section{Acknowledgements}

This research has made use of data, software and/or web tools obtained from the Gravitational Wave Open Science Center (https://www.gw-openscience.org/), a service of LIGO Laboratory, the LIGO Scientific Collaboration and the Virgo Collaboration. LIGO Laboratory and Advanced LIGO are funded by the United States National Science Foundation (NSF) as well as the Science and Technology Facilities Council (STFC) of the United Kingdom, the Max-Planck-Society (MPS), and the State of Niedersachsen/Germany for support of the construction of Advanced LIGO and construction and operation of the GEO600 detector. Additional support for Advanced LIGO was provided by the Australian Research Council. Virgo is funded, through the European Gravitational Observatory (EGO), by the French Centre National de Recherche Scientifique (CNRS), the Italian Istituto Nazionale di Fisica Nucleare (INFN) and the Dutch Nikhef, with contributions by institutions from Belgium, Germany, Greece, Hungary, Ireland, Japan, Monaco, Poland, Portugal, Spain.

The authors gratefully acknowledge the support of the United States
National Science Foundation (NSF) for the construction and operation of the
LIGO Laboratory and Advanced LIGO as well as the Science and Technology Facilities Council (STFC) of the
United Kingdom, the Max-Planck-Society (MPS), and the State of
Niedersachsen/Germany for support of the construction of Advanced LIGO 
and construction and operation of the GEO600 detector. 
Additional support for Advanced LIGO was provided by the Australian Research Council.
The authors gratefully acknowledge the Italian Istituto Nazionale di Fisica Nucleare (INFN),  
the French Centre National de la Recherche Scientifique (CNRS) and
the Foundation for Fundamental Research on Matter supported by the Netherlands Organisation for Scientific Research, 
for the construction and operation of the Virgo detector
and the creation and support  of the EGO consortium. 
The authors also gratefully acknowledge research support from these agencies as well as by 
the Council of Scientific and Industrial Research of India, 
the Department of Science and Technology, India,
the Science \& Engineering Research Board (SERB), India,
the Ministry of Human Resource Development, India,
the Spanish  Agencia Estatal de Investigaci\'on,
the Vicepresid\`encia i Conselleria d'Innovaci\'o, Recerca i Turisme and the Conselleria d'Educaci\'o i Universitat del Govern de les Illes Balears,
the Conselleria d'Educaci\'o, Investigaci\'o, Cultura i Esport de la Generalitat Valenciana,
the National Science Centre of Poland,
the Swiss National Science Foundation (SNSF),
the Russian Foundation for Basic Research, 
the Russian Science Foundation,
the European Commission,
the European Regional Development Funds (ERDF),
the Royal Society, 
the Scottish Funding Council, 
the Scottish Universities Physics Alliance, 
the Hungarian Scientific Research Fund (OTKA),
the Lyon Institute of Origins (LIO),
the Paris \^{I}le-de-France Region, 
the National Research, Development and Innovation Office Hungary (NKFIH), 
the National Research Foundation of Korea,
Industry Canada and the Province of Ontario through the Ministry of Economic Development and Innovation, 
the Natural Science and Engineering Research Council Canada,
the Canadian Institute for Advanced Research,
the Brazilian Ministry of Science, Technology, Innovations, and Communications,
the International Center for Theoretical Physics South American Institute for Fundamental Research (ICTP-SAIFR), 
the Research Grants Council of Hong Kong,
the National Natural Science Foundation of China (NSFC),
the Leverhulme Trust, 
the Research Corporation, 
the Ministry of Science and Technology (MOST), Taiwan
and
the Kavli Foundation.
The authors gratefully acknowledge the support of the NSF, STFC, INFN and CNRS for provision of computational resources.

\section{Competing interests}

No conflict of interest

\end{document}